\documentclass[twoside,leqno]{article}
\usepackage{amsmath}
\usepackage{amsthm}
\usepackage{amssymb}
\usepackage{amscd}
\usepackage{graphicx}
\usepackage{epsfig}

\usepackage{latexsym}
\usepackage{amsfonts}
\input xy
\xyoption{all} \tolerance=500

-\headheight\advance\topmargin by -\headsep

\numberwithin{equation}{section} \allowdisplaybreaks

\newtheorem{theorem}{Theorem}[section]

\theoremstyle{definition}

\begin{document}
\font\black=cmbx10 \font\sblack=cmbx7 \font\ssblack=cmbx5
\font\blackital=cmmib10  \skewchar\blackital='177
\font\sblackital=cmmib7 \skewchar\sblackital='177
\font\ssblackital=cmmib5 \skewchar\ssblackital='177
\font\sanss=cmss10 \font\ssanss=cmss8 
\font\sssanss=cmss8 scaled 600 \font\blackboard=msbm10
\font\sblackboard=msbm7 \font\ssblackboard=msbm5
\font\caligr=eusm10 \font\scaligr=eusm7 \font\sscaligr=eusm5
\font\blackcal=eusb10 \font\fraktur=eufm10 \font\sfraktur=eufm7
\font\ssfraktur=eufm5 \font\blackfrak=eufb10

\font\bsymb=cmsy10 scaled\magstep2
\def\all#1{\setbox0=\hbox{\lower1.5pt\hbox{\bsymb
       \char"38}}\setbox1=\hbox{$_{#1}$} \box0\lower2pt\box1\;}
\def\exi#1{\setbox0=\hbox{\lower1.5pt\hbox{\bsymb \char"39}}
       \setbox1=\hbox{$_{#1}$} \box0\lower2pt\box1\;}

\def\mi#1{{\fam1\relax#1}}
\def\tx#1{{\fam0\relax#1}}

\newfam\bifam
\textfont\bifam=\blackital \scriptfont\bifam=\sblackital
\scriptscriptfont\bifam=\ssblackital
\def\bi#1{{\fam\bifam\relax#1}}

\newfam\blfam
\textfont\blfam=\black \scriptfont\blfam=\sblack
\scriptscriptfont\blfam=\ssblack
\def\rbl#1{{\fam\blfam\relax#1}}

\newfam\bbfam
\textfont\bbfam=\blackboard \scriptfont\bbfam=\sblackboard
\scriptscriptfont\bbfam=\ssblackboard
\def\bb#1{{\fam\bbfam\relax#1}}

\newfam\ssfam
\textfont\ssfam=\sanss \scriptfont\ssfam=\ssanss
\scriptscriptfont\ssfam=\sssanss
\def\sss#1{{\fam\ssfam\relax#1}}

\newfam\clfam
\textfont\clfam=\caligr \scriptfont\clfam=\scaligr
\scriptscriptfont\clfam=\sscaligr
\def\cl#1{{\fam\clfam\relax#1}}

\newfam\frfam
\textfont\frfam=\fraktur \scriptfont\frfam=\sfraktur
\scriptscriptfont\frfam=\ssfraktur
\def\fr#1{{\fam\frfam\relax#1}}

\def\cb#1{\hbox{$\fam\gpfam\relax#1\textfont\gpfam=\blackcal$}}

\def\hpb#1{\setbox0=\hbox{${#1}$}
    \copy0 \kern-\wd0 \kern.2pt \box0}
\def\vpb#1{\setbox0=\hbox{${#1}$}
    \copy0 \kern-\wd0 \raise.08pt \box0}

\def\pmb#1{\setbox0\hbox{${#1}$} \copy0 \kern-\wd0 \kern.2pt \box0}
\def\pmbb#1{\setbox0\hbox{${#1}$} \copy0 \kern-\wd0
      \kern.2pt \copy0 \kern-\wd0 \kern.2pt \box0}
\def\pmbbb#1{\setbox0\hbox{${#1}$} \copy0 \kern-\wd0
      \kern.2pt \copy0 \kern-\wd0 \kern.2pt
    \copy0 \kern-\wd0 \kern.2pt \box0}
\def\pmxb#1{\setbox0\hbox{${#1}$} \copy0 \kern-\wd0
      \kern.2pt \copy0 \kern-\wd0 \kern.2pt
      \copy0 \kern-\wd0 \kern.2pt \copy0 \kern-\wd0 \kern.2pt \box0}
\def\pmxbb#1{\setbox0\hbox{${#1}$} \copy0 \kern-\wd0 \kern.2pt
      \copy0 \kern-\wd0 \kern.2pt
      \copy0 \kern-\wd0 \kern.2pt \copy0 \kern-\wd0 \kern.2pt
      \copy0 \kern-\wd0 \kern.2pt \box0}

\def\cdotss{\mathinner{\cdotp\cdotp\cdotp\cdotp\cdotp\cdotp\cdotp
        \cdotp\cdotp\cdotp\cdotp\cdotp\cdotp\cdotp\cdotp\cdotp\cdotp
        \cdotp\cdotp\cdotp\cdotp\cdotp\cdotp\cdotp\cdotp\cdotp\cdotp
        \cdotp\cdotp\cdotp\cdotp\cdotp\cdotp\cdotp\cdotp\cdotp\cdotp}}

\font\frak=eufm10 scaled\magstep1 \font\fak=eufm10 scaled\magstep2
\font\fk=eufm10 scaled\magstep3 \font\scriptfrak=eufm10
\font\tenfrak=eufm10


\mathchardef\za="710B  
\mathchardef\zb="710C  
\mathchardef\zg="710D  
\mathchardef\zd="710E  
\mathchardef\zve="710F 
\mathchardef\zz="7110  
\mathchardef\zh="7111  
\mathchardef\zvy="7112 
\mathchardef\zi="7113  
\mathchardef\zk="7114  
\mathchardef\zl="7115  
\mathchardef\zm="7116  
\mathchardef\zn="7117  
\mathchardef\zx="7118  
\mathchardef\zp="7119  
\mathchardef\zr="711A  
\mathchardef\zs="711B  
\mathchardef\zt="711C  
\mathchardef\zu="711D  
\mathchardef\zvf="711E 
\mathchardef\zq="711F  
\mathchardef\zc="7120  
\mathchardef\zw="7121  
\mathchardef\ze="7122  
\mathchardef\zy="7123  
\mathchardef\zf="7124  
\mathchardef\zvr="7125 
\mathchardef\zvs="7126 
\mathchardef\zf="7127  
\mathchardef\zG="7000  
\mathchardef\zD="7001  
\mathchardef\zY="7002  
\mathchardef\zL="7003  
\mathchardef\zX="7004  
\mathchardef\zP="7005  
\mathchardef\zS="7006  
\mathchardef\zU="7007  
\mathchardef\zF="7008  
\mathchardef\zW="700A  

\newcommand{\be}{\begin{equation}}
\newcommand{\ee}{\end{equation}}
\newcommand{\ra}{\rightarrow}
\newcommand{\lra}{\longrightarrow}
\newcommand{\bea}{\begin{eqnarray}}
\newcommand{\eea}{\end{eqnarray}}
\newcommand{\beas}{\begin{eqnarray*}}
\newcommand{\eeas}{\end{eqnarray*}}
\def\*{{\textstyle *}}
\newcommand{\R}{{\mathbb R}}
\newcommand{\T}{{\mathbb T}}
\newcommand{\C}{{\mathbb C}}
\newcommand{\unit}{{\mathbf 1}}
\newcommand{\SL}{SL(2,\C)}
\newcommand{\Sl}{sl(2,\C)}
\newcommand{\SU}{SU(2)}
\newcommand{\su}{su(2)}
\def\ssT{\sss T}
\newcommand{\G}{{\goth g}}
\newcommand{\D}{{\rm d}}
\newcommand{\Df}{{\rm d}^\zF}
\newcommand{\de}{\,{\stackrel{\rm def}{=}}\,}
\newcommand{\we}{\wedge}
\newcommand{\nn}{\nonumber}
\newcommand{\ot}{\otimes}
\newcommand{\s}{{\textstyle *}}
\newcommand{\ts}{T^\s}
\newcommand{\oX}{\stackrel{o}{X}}
\newcommand{\oD}{\stackrel{o}{D}}
\newcommand{\obD}{\stackrel{o}{\bD}}
\newcommand{\pa}{\partial}
\newcommand{\ti}{\times}
\newcommand{\A}{{\cal A}}
\newcommand{\Li}{{\cal L}}
\newcommand{\ka}{\mathbb{K}}
\newcommand{\find}{\mid}
\newcommand{\ad}{{\rm ad}}
\newcommand{\rS}{]^{SN}}
\newcommand{\rb}{\}_P}
\newcommand{\p}{{\sf P}}
\newcommand{\h}{{\sf H}}
\newcommand{\X}{{\cal X}}
\newcommand{\I}{\,{\rm i}\,}
\newcommand{\rB}{]_P}
\newcommand{\Ll}{{\pounds}}
\def\lna{\lbrack\! \lbrack}
\def\rna{\rbrack\! \rbrack}
\def\rnaf{\rbrack\! \rbrack_\zF}
\def\rnah{\rbrack\! \rbrack\,\hat{}}
\def\lbo{{\lbrack\!\!\lbrack}}
\def\rbo{{\rbrack\!\!\rbrack}}
\def\lan{\langle}
\def\ran{\rangle}
\def\zT{{\cal T}}
\def\tU{\tilde U}
\def\ati{{\stackrel{a}{\times}}}
\def\sti{{\stackrel{sv}{\times}}}
\def\aot{{\stackrel{a}{\ot}}}
\def\sati{{\stackrel{sa}{\times}}}
\def\saop{{\stackrel{sa}{\op}}}
\def\bwa{{\stackrel{a}{\bigwedge}}}
\def\svop{{\stackrel{sv}{\oplus}}}
\def\saot{{\stackrel{sa}{\otimes}}}
\def\cti{{\stackrel{cv}{\times}}}
\def\cop{{\stackrel{cv}{\oplus}}}
\def\dra{{\stackrel{\xd}{\ra}}}
\def\bdra{{\stackrel{\bd}{\ra}}}
\def\bAff{\mathbf{Aff}}
\def\Aff{\sss{Aff}}
\def\bHom{\mathbf{Hom}}
\def\Hom{\sss{Hom}}
\def\bt{{\boxtimes}}
\def\sot{{\stackrel{sa}{\ot}}}
\def\bp{{\boxplus}}
\def\op{\oplus}
\def\bwak{{\stackrel{a}{\bigwedge}\!{}^k}}
\def\aop{{\stackrel{a}{\oplus}}}
\def\ix{\operatorname{i}}
\def\V{{\cal V}}
\def\cD{{\cal D}}
\def\cC{{\cal C}}
\def\cE{{\cal E}}
\def\cL{{\cal L}}
\def\cN{{\cal N}}
\def\cR{{\cal R}}
\def\cJ{{\cal J}}
\def\cT{{\cal T}}
\def\cH{{\cal H}}
\def\cS{{\cal S}}
\def\bA{\mathbf{A}}
\def\bI{\mathbf{I}}
\def\wh{\widehat}
\def\wt{\widetilde}
\def\ol{\overline}
\def\ul{\underline}
\def\Sec{\sss{Sec}}
\def\Lin{\sss{Lin}}
\def\ader{\sss{ADer}}
\def\ado{\sss{ADO^1}}
\def\adoo{\sss{ADO^0}}
\def\AS{\sss{AS}}
\def\bAS{\sss{AS}}
\def\bLS{\sss{LS}}
\def\bAP{\sss{AV}}
\def\bLP{\sss{LP}}
\def\AP{\sss{AP}}
\def\LP{\sss{LP}}
\def\LS{\sss{LS}}
\def\Z{\mathbf{Z}}
\def\oZ{\overline{\bZ}}
\def\oA{\overline{\bA}}
\def\cim{{C^\infty(M)}}
\def\de{{\cal D}^1}
\def\la{\langle}
\def\ran{\rangle}
\def\<{\langle}
\def\>{\rangle}
\def\bcS{\mathbb S}
\def\by{{\bi y}}
\def\bs{{\bi s}}
\def\bc{{\bi c}}
\def\bd{{\bi d}}
\def\bh{{\bi h}}
\def\bD{{\bi D}}
\def\bY{{\bi Y}}
\def\bX{{\bi X}}
\def\bL{{\bi L}}
\def\bV{{\bi V}}
\def\bW{{\bi W}}
\def\bS{{\bi S}}
\def\bT{{\bi T}}
\def\bC{{\bi C}}
\def\bE{{\bi E}}
\def\bF{{\bi F}}
\def\bP{{\bi P}}
\def\bp{{\bi p}}
\def\bz{{\bi z}}
\def\bZ{{\bi Z}}
\def\bq{{\bi q}}
\def\bQ{{\bi Q}}
\def\bx{{\bi x}}

\def\sA{{\sss A}}
\def\sC{{\sss C}}
\def\sD{{\sss D}}
\def\sG{{\sss G}}
\def\sH{{\sss H}}
\def\sI{{\sss I}}
\def\sJ{{\sss J}}
\def\sK{{\sss K}}
\def\sL{{\sss L}}
\def\sO{{\sss O}}
\def\sP{{\sss P}}
\def\sPh{{\sss P\sss h}}
\def\sT{{\sss T}}
\def\sV{{\sss V}}
\def\sR{{\sss R}}
\def\sS{{\sss S}}
\def\sE{{\sss E}}
\def\sF{{\sss F}}
\def\st{{\sss t}}
\def\sg{{\sss g}}
\def\sx{{\sss x}}
\def\sv{{\sss v}}
\def\sw{{\sss w}}
\def\sQ{{\sss Q}}
\def\sj{{\sss j}}
\def\sq{{\sss q}}
\def\xa{\tx{a}}
\def\xc{\tx{c}}
\def\xd{\tx{d}}
\def\xi{\tx{i}}
\def\xD{\tx{D}}
\def\xV{\tx{V}}
\def\xF{\tx{F}}


\setcounter{page}{1} \thispagestyle{empty}
\bigskip

\bigskip

\title{The Schr\"odinger operator in Newtonian space-time\thanks{Research supported by the Polish Ministry of
Scientific Research and Information Technology under the grant No.
2 P03A 036 25.}}

        \author{
        Katarzyna  Grabowska$^1$, Janusz Grabowski$^2$, Pawe\l\ Urba\'nski$^1$\\
        \\
         $^1$ {\it Physics Department}\\
                {\it University of Warsaw} \\
         $^2$ {\it Institute of Mathematics}\\
                {\it Polish Academy of Sciences}
                }
\date{}
\maketitle
\begin{abstract}
The Schr\"odinger operator on the Newtonian space-time is defined
in a way which is independent on the class of inertial observers.
In this picture the Schr\"odinger operator acts not on functions
on the space-time but on sections of certain one-dimensional
complex vector bundle over space-time. This bundle, constructed
from the data provided by all possible inertial observers, has no
canonical trivialization, so these sections cannot be viewed as
functions on the space-time. The presented framework is
conceptually four-dimensional and does not involve any {\em ad
hoc} or axiomatically introduced geometrical structures. It is
based only on the traditional understanding of the Schr\"odinger
operator in a given reference frame and it turns out to be
strictly related to the frame-independent formulation of
analytical Newtonian mechanics that makes a bridge between the
classical and quantum theory.


\end{abstract}
\section{Introduction}

In the papers \cite{GGU1,GGU2,GGU3,U} we have presented an
approach to differential geometry in which sections of a
one-dimensional affine bundle over a manifold have been used
instead of functions on the manifold. This approach, initiated by
W.~M.~Tulczyjew in \cite{TU,TUZ}, has been successfully applied to
frame-independent description of different systems, in particular
to a frame-independent formulation of Newtonian mechanics
\cite{GU}.

The latter problem is closely related to the problem of
frame-independent formulation of wave mechanics in the Newtonian
space-time. It is known that a solution of the Schr\"odinger
equation in one inertial frame will not, in general, satisfy the
Schr\"odinger equation in a different frame. The same quantum
state of a particle must be represented by a different wave
function in reference to a different inertial frame. The
corresponding gauge transformation of solutions of the
Schr\"odinger equation was known already to W.~Pauli \cite{Pa}.
Many ways of solving this problem have been proposed in the
literature. For instance, a general axiomatic theory of quantum
bundles, quantum metrics, quantum connections etc. has been
developed in \cite{JM} to deal with a covariant description of
Schr\"odinger operators in curved space-times. Another general
fibre bundle formulation of nonrelativistic quantum mechanics has
been proposed in a series of papers \cite{Il}.

An approach which is the closest to what we propose in this paper
is a frame-independent formulation of wave mechanics by extending
Newtonian space-time to five-dimensional Galilei space
\cite{DB,WMT2,WMT3}. The corresponding geometry is associated with
the Bargmann group - nontrivially extended Galilei group
\cite{Ba}.

In the present paper we change this view-point a little bit,
making the whole theory four-dimensional again. For simplicity, we
deal with the flat Newtonian space-time and the very standard
Schr\"odinger operator to show that a frame-independent
formulation of wave mechanics is possible in terms of a
canonically constructed line bundle. In this picture the
Schr\"odinger operator acts not on functions on the space-time but
on sections of certain one-dimensional complex vector bundle over
space-time. This bundle, constructed from the data provided by all
possible inertial observers, has no canonical trivialization, so
these sections cannot be viewed as functions on the space-time.

We want to stress three facts. First, we do not look just for
transformations rules for solutions of the Schr\"odinger equation
in different reference frames, but we built the bundle whose
sections represent the arguments of the Schr\"odinger operator and
we give to the operator itself a covariant geometric meaning.

Second, we show that the transformation rules, so the bundle, are
unique up to unavoidable change by constant factors - the
integration is always slightly non-unique.

And last but not least, we prove that the proposed formulation is
strictly related to the frame-independent formulation of
analytical Newtonian mechanics \cite{GU}. This makes a bridge
between the classical and quantum theory which, in our opinion, is
not understood completely yet and usually not present in the
literature.

\section{Newtonian space-time}

The {\it Newtonian space-time} (some authors prefer to call it
{\em Galilean space-time}, but we follow the terminology of
Benenti \cite{Be} and Tulczyjew \cite{WMT2}) is a system $(N,\tau,
g)$, where $N$ is a four-dimensio\-nal affine space for which, say
$V$, is the model vector space, where $\tau$ is a non-zero element
of $V^\ast$, and where $g\colon E_0\rightarrow E_0^\ast$
represents an Euclidean metric on $E_0=\ker\tau$. The
corresponding scalar product reads $\< v\mid v'\>=(g(v))(v')$ and
the corresponding norm $\Vert v\Vert=\sqrt{\< v\mid v\>}$. The
elements of the space $N$ represent events. The time elapsed
between two events is measured by $\tau$:
    $$\Delta t(x,x')=\tau(x-x')$$
and the distance between two simultaneous events is measured by
$g$:
    $$d(x,x')=\Vert x-x'\Vert.$$
    The space-time $N$ is fibred over the time $T=N\slash E_0 $
which is a one-dimensional affine space modelled on $\R$.

Let $E_1$ be an affine subspace of $V$ defined by the equation
$\tau(v)=1$. The model vector space for this subspace is $E_0$. An
element of $E_1$ represents velocity of a particle. The affine
structure of $N$ allows us to associate to an element $u$ of $E_1$
the family of inertial observers that move in the space-time with
the constant velocity $u$. In this way we can interpret an element
of $E_1$ also as a class of inertial reference frames. For a fixed
inertial frame  $u\in E_1$ and a point $x_0\in N$, we can identify
$N$ with $E_0\ti\R$ by
\be \zF_{(x_0,u)}:N\ra E_0\ti\R,\quad x\mapsto \left((x-x_0)-
\zt(x-x_0)u,\zt(x-x_0)\right).\ee A change of the inertial
reference frame results in the change of this identification and
it is represented by
\bea\label{1}\zY^{(x_0',u')}_{(x_0,u)}&=&\zF_{(x'_0,u')}\circ\zF_{(x_0,u)}^{-1}:
E_0\ti\R\ra E_0\ti\R, \\ (v,t)&\mapsto &
(v-((x'_0-x_0)-\zt(x'_0-x_0)u')-(u'-u)t,t-\zt(x'_0-x_0)).\eea We
can fix linear coordinates $y=(y_i):E_0\ra\R^3$ in $E_0$ so that
$\Vert v\Vert^2=\sum_iy_i^2(v)$. Then, with every inertial frame
$(x_0,u)$, we can associate coordinates $(y,t)$ in $N$, thus $V$,
with
$(y,t)(x)=\zf_{(x_0,u)}(x)=(y(x-x_0-\zt(x-x_0)u),\zt(x-x_0))$, and
the change of coordinates is
\be\label{2}\zY^{(x_0',u')}_{(x_0,u)}(y,t)=(y-\zD y-y(v)t,t-\zD t),
\ee
where $(\zD y,\zD t)\in\R^3\ti\R$ are coordinates of $x'_0-x_0\in
V$ for the observer $(x_0,u)$ and $y(v)\in\R^3$ are coordinates of
$v=u'-u\in E_0$.

\section{The Schr\"odinger operator}
The classical Schr\"odinger operator for a particle of mass $m$
and a potential $\wt{U}\in C^\infty(\R^3\ti\R)$ is a second order
complex differential operator which, in coordinates described
above, reads
\be\label{S} \cS_{\wt{U}}^m\zc=i\hbar\frac{\pa\zc}{\pa
t}+\frac{\hbar^2}{2m}\sum_k\frac{\pa^2\zc}{\pa y_i^2}-\wt{U}\zc.
\ee Here, $\sum_k\frac{\pa^2}{\pa y_i^2}$ is clearly the spatial
Laplace-Beltrami operator associated with the metric $g$. The
problem is that, if assumed as acting on functions, the
Schr\"odinger operator (\ref{S}) is not invariant with respect to
the change of coordinates (\ref{2}) associated with the choice of
another inertial frame. On the other hand, by  arguments coming
from physics, the form of the Schr\"odinger operator should be
independent on the choice of inertial observer.

The solution we propose is that the Schr\"odinger operator acts in
fact on sections of certain 1-dimensional complex vector bundle
$S_m$ over $N$ (we will call it {\em Schr\"odinger bundle}) which
is trivializable but with no canonical trivialization. Thus the
situation is completely parallel to the one we encounter for
frame-independent description of the standard lagrangian in
Newtonian mechanics \cite{GU}.

Since the bundle $S_m$ has no canonical trivialization, we have to
combine every change of coordinates (\ref{2}) in $N$ with a linear
change in values of wave functions. However, we can simplify this
problem a little bit. Since, as can be easily seen, the part
corresponding to the potential $\wt{U}$ associated with a function
$U$ on $N$ behaves properly and the Schr\"odinger operator is
invariant with respect to the change of coordinates associated
with observers moving with the same velocity, $u=u'$, we can
assume that $\wt{U}=0$ and $x'_0=x_0$, so we reduce
transformations to vertical (spatial) ones. Thus we shall look for
an action of the commutative group $E_0$ in $\R^3\ti\R\ti\C$ of
the form
\be\label{4}\Psi_v(y_k,t,z)= \left(y_k-y_k(v)t,t,e^{F_v(y,t)}z\right),\ee
corresponding to the representation of $E_0$ in the algebra
$C^\infty_\C(\R^3\ti\R)$ of complex-valued functions on
$\R^3\ti\R$,
\be\label{4a}T_v(\zc)(y,t)=e^{F_v(y,t)}\zc(y-y(v)t,t),\ee such that
the "free" Schr\"odinger operator
\be\label{S1}\cS_0^m\zc=i\hbar\frac{\pa\zc}{\pa
t}+\frac{\hbar^2}{2m}\sum_k\frac{\pa^2\zc}{\pa y_i^2}.\ee remains
unchanged:
\be\label{5} \cS_0^m\left(e^{F_v(y,t)}\zc(y-y(v)t,t)\right)=
e^{F_v(y,t)}\cS_0^m(\zc)(y-y(v)t,t).\ee

\medskip\noindent
{\bf Remark.} That our spatial part is 3-dimensional is motivated
by physics. However, from the mathematical point of view, there is
no difference if we use other dimensions. All considerations and
proofs remain unchanged if we use $\R^n\ti\R\ti\C$ instead of
$\R^3\ti\R\ti\C$.

\section{The explicit transformations}
Let us look what the function $F_v$ could be in order that
(\ref{5}) is satisfied. Straightforward calculations show that
(\ref{5}) is equivalent to \bea
&\zc(y-y(v)t,t)\left(i(\pa_tF)(y,t)+\frac{\hbar}{2m}\left(
\sum_k(\pa_{y_k}F)^2(y,t)+\sum_k(\pa_{y_k}^2F)(y,t)\right)\right)+\\
& \sum_k(\pa_{y_k}\zc)(y-y(v)t,t)\left(\frac{\hbar}{m}
(\pa_{y_k}F)(y,t)-iy_k(v)\right)=0\nn\eea for all complex
functions $\zc$ on $\R^3\ti\R$. Since $\zc$ is arbitrary, this, in
turn, is equivalent to the system of equations
\bea\label{r1} &i(\pa_tF)(y,t)+\frac{\hbar}{2m}\left(\sum_k(\pa_{y_k}F)^2(y,t)+
\sum_k(\pa_{y_k}^2F)(y,t)\right)=0,\\
&\frac{\hbar}{m}(\pa_{y_k}F)(y,t)-iy_k(v)=0\,,k=1,2,3\,.\label{r2}
\eea From (\ref{r2}) it follows that $\pa_{y_k}^2F=0$, $k=1,2,3$,
so that (\ref{r1}) reduces to
\be\label{r1a}i(\pa_tF)(y,t)-\frac{m}{2\hbar}\sum_ky_k^2(v)=0. \ee
The equations (\ref{r2}) and (\ref{r1a}) for partial derivatives
determine $F$ up to a constant, so, as can be easily seen,
\be\label{F}F_v(y,t)=\frac{im}{\hbar}\left(
\sum_ky_k(v)y_k-\frac{t}{2}\sum_ky_k^2(v)\right)+c\,. \ee But
$T_v(\zc)(y,t)=e^{F_v(y,t)}\zc(y-y(v)t,t)$ must be a
representation, i.e.
\be\label{rep}T_v\circ T_{v'}=T_{v+v'}.\ee Direct calculations
show that (\ref{rep}) is satisfied with $F_v$ as in (\ref{F}) if
and only if $c=0$. In this way we get the following.

\begin{theorem}\label{t1} The map $\R^3\ni v\mapsto T_v$, where $T_v$ is a
linear operator in $C^\infty_\C(\R^3\ti\R)$ defined by
\be\label{rep1} T_v(\zc)(y,t)=\exp{\left(\frac{im}{\hbar}\left(
\sum_ky_k(v)y_k-\frac{t}{2}\sum_ky_k^2(v)\right)\right)}\zc(y-y(v)t,t),\ee
is a representation of $\R^3$ in $C^\infty_\C(\R^3\ti\R)$ leaving
invariant the Schr\"odinger operator (\ref{S1}).
\end{theorem}
The fact that the above transformations act on solutions of the
Schr\"odinger equation in different reference frames is known (see
e.g. \cite[p. 100]{Pa}). Here, indepentently, we have found these
transformations in order to recognize properly the arguments of
the Schr\"odinger operator and the operator itself, and we have
proved the uniqueness.

\section{The Schr\"odinger bundle}
Let us now fix $x_0\in N$. In the trivial 1-dimensional complex
bundle $E_1\ti N\ti\C$ over $E_1\ti N$ we consider the following
linear action of the additive group $E_0$:
\begin{equation}
R_v(u,x,z)=\left(u+v,x,z\cdot\exp{\left[\frac{m}{i\hbar}\left(\<v\mid
x-x_0-\zt(x-x_0)u\>- \frac{\zt(x-x_0)}{2}\Vert
v\Vert^2\right)\right]}\right).
\end{equation}
This is indeed an action, since
\beas &\<v+v'\mid x-x_0-\zt(x-x_0)(u+v)\>-
\frac{\zt(x-x_0)}{2}\Vert v+v'\Vert^2=\\
&\<v\mid x-x_0-\zt(x-x_0)u\>- \frac{\zt(x-x_0)}{2}\Vert v\Vert^2+
\<v'\mid x-x_0-\zt(x-x_0)(u+v)\>- \frac{\zt(x-x_0)}{2}\Vert
v'\Vert^2.
\eeas
Since the action is free and linear and since the orbits project
onto $E_1\ti\{x\}$, the quotient space, i.e. the family of orbits,
forms a new vector bundle $S_m$ over $N$. If $[u,x,z]$ is the
class (orbit) through $(u,x,z)$, the addition in the bundle is
represented by
$$[u,x,z]+[u+v,x,z']=[u,x,z+z'\cdot\exp{\left[\frac{im}{\hbar}\left(\<v\mid x-x_0-\zt(x-x_0)u\>-
\frac{\zt(x-x_0)}{2}\Vert v\Vert^2\right)\right]}
$$
and the multiplication by complex numbers $\za$ is given by
$$\za[u,x,z]=[u,x,\za\cdot z].$$
This bundle we shall call the {\em Schr\"odinger bundle}
associated with the mass $m$. It is clear that the bundle is
trivializable, since, for fixed $u$, the global section
$\zs_u(x)=[u,x,1]$ is nowhere-vanishing. On the other hand, there
is no canonical trivialization, since the choice of $u$ is
arbitrary.

We have a system of global trivializations $\Psi_u:
S_m\ra\R^3\ti\R\ti\C$, $u\in E_1$,
\beas &\Psi_u([u+v,x,z])=\left(\zf_u(x),z\cdot\exp{\left[\frac{m}{i\hbar}
\left(\<v\mid x-x_0-\zt(x-x_0)u\>-
\frac{\zt(x-x_0)}{2}\Vert v\Vert^2\right)\right]}\right)=\\
&\left(y(x-x_0-\tau(x-x_0)u),\tau(x-x_0),
z\cdot\exp{\left[\frac{m}{i\hbar}\left(\<v\mid
x-x_0-\zt(x-x_0)u\>- \frac{\zt(x-x_0)}{2}\Vert
v\Vert^2\right)\right]}\right).
\eeas
Note first that $\Psi_u$ is well defined, i.e. does not depend on
the representant $(u+v,x,z)$ in the class. Moreover, the change of
trivializations in coordinates reads
$$(\Psi_{u+v}\circ\Psi_u^{-1})(y,t,z)=\left(y-y(v)t,t,e^{F_v(y,t)}z\right)$$
with
$$F_v(y,t)=\frac{im}{\hbar}\left(
\sum_ky_k(v)y_k-\frac{t}{2}\sum_ky_k^2(v)\right).$$ According to
Theorem \ref{t1}, the differential operator $\bcS_u^m$ on $S_m$
that corresponds to $\cS_0^m$ on the trivial 1-dimensional vector
bundle $\R^3\ti\R\ti\C$, does not depend on the trivialization
$\Psi_u$, so it gives rise to a well-defined differential operator
$\bcS_0^m$ on $S_m$. Choosing a potential $U\in C^\infty_\C(N)$ we
can write the full Schr\"odinger operator as
$\bcS_U^m\zc=\bcS_0^m\zc+U\zc$ acting on sections of $S_m$. It
corresponds, {\em via} $\Psi_u$, to
$\cS_{\wt{U}}^m=\cS_0^m+\wt{U}$ with
$\wt{U}=U\circ\zf_{(x_0,u)}^{-1}$.  Moreover, the bundle
isomorphisms $\Psi_{u+v}\circ\Psi_u^{-1}$ relate
$\cS_0^m+U\circ\zf_{(x_0,u)}^{-1}$ with
$\cS_0^m+U\circ\zf_{(x_0,u+v)}^{-1}$. We can summarize these
observations as follows.
\begin{theorem}
For any function $U$ on the newtonian space-time $N$ (potential)
and every $m\in\R$ (mass) there is a well-defined differential
operator $\bcS_U^m$ ({\em the Schr\"odinger operator}), acting on
sections of the Schr\"odinger bundle $S_m$. This operator
corresponds, {\em via} the trivialization $\Psi_u$, to the
differential operator
\be\label{Ss} \cS_U^m\zc=i\hbar\frac{\pa\zc}{\pa
t}+\frac{\hbar^2}{2m}\sum_k\frac{\pa^2\zc}{\pa
y_i^2}-(U\circ\zf_u^{-1})\zc
\ee
acting on complex functions $\zc(y,t)$ on $\R^3\ti\R$.
\end{theorem}

\section{Relation to Newtonian mechanics}
In addition to the vector bundle operation in the Schr\"odinger
bundle $S_m$ we can define a norm in its fibers by
    $$ \| [u,x,z]\| = |z| .$$
Consequently, we can consider the bundle $S_m$ as associated with
a principal $U(1)$-bundle $B$. It is a subbundle of normalized
vectors in  $S_m$. By means of a group homomorphism
     $$ \R  \rightarrow U(1)\colon  r \mapsto \exp\left(\frac{i r}{\hbar}\right), $$
     the bundle $B$ can be considered as the reduced $(\R,+)$-bundle $Z_m$. It is an AV-bundle
     in terminology of  \cite{GGU2}.
An element of $Z_m$ is an equivalence class of triples $(u,x,r)
\in E_1\ti N \ti \R$, where two triples $(u,x,r)$ and
$(u',x',r')$ are equivalent if $x=x'$ and
    $$r' = r +m\left(\frac{\zt(x-x_0)}{2}\Vert u' - u\Vert^2- \langle x-x_0- \zt(x-x_0) u\mid u'-u\rangle\right) .$$
    The corresponding affine covector  (see \cite{GGU2}) is an equivalence class of triples
    $(u,x,p) \in E_1\ti N \ti V^\*$. Two such triples $(u,x,p)$ and $(u',x',p')$  are
    equivalent if $x=x'$ and
 \be p'= p +m\cdot\xd\left(\frac{\zt(x-x_0)}{2}
 \Vert u' - u\Vert^2- \langle x-x_0- \zt(x-x_0)u\mid u'-u\rangle
 \right).\ee
Since
\beas&\xd\left(\frac{\zt(x-x_0)}{2}
 \Vert u' - u\Vert^2- \langle x-x_0- \zt(x-x_0)u\mid u'-u\rangle
 \right)(v)=\\
 &\frac{1}{2}
 \Vert u' - u\Vert^2\zt(v)- \langle v- \zt(v)u\mid
 u'-u\rangle,\eeas
 we get
 $$p'=p+m\zs(u',u),$$
 where
 $$\zs(u',u)(v)=\la u'-u\mid v-\zt(v)\frac{u'+u}{2}\ran,$$
which is precisely the relation used in \cite{GU} to define the
affine phase space for a Newtonian particle of mass $m$.
\section{Concluding remarks}
We have proposed an understanding of the Schr\"odinger operator on
the Newtonian space-time in a way which makes it independent on
the class of inertial observers. In this picture the Schr\"odinger
operator acts not on functions on the space-time but on sections
of certain one-dimensional complex vector bundle over space-time.
This bundle, constructed from the data provided by all possible
inertial observers, has no canonical trivialization, so these
sections cannot be viewed as functions on the space-time. The
presented framework is conceptually four-dimensional (the base is
the traditional Newtonian space-time but the values of wave
functions are not true numbers) and strictly related to the
frame-independent formulation of analytical Newtonian mechanics
that makes a bridge between the classical and quantum theory.

What we propose can look for the first sight like a formal
modification of the five-dimensional approach, but it is in fact a
real conceptual change. We insist on using the picture in which
wave functions live on the standard space-time, since, in our
opinion, finding appropriate geometric structures for physics is
not just a pure mathematical game of formal manipulations. It
might result in correcting our understanding of the physical
content of the notions used in the theory. It is sufficient to
mention the impact on physics of the Einstein's proper geometric
formulation of what is space, time, and matter. Moreover, our
approach does not involve any {\em ad hoc} or axiomatically
introduced geometrical structures and is based only on the
traditional understanding of the Schr\"odinger operator in a given
reference frame. This makes it mathematically simple,
demonstrative, and respecting the postulate of Occam's Raizor.

To finish, let us mention that there is a way to understand the
Schr\"odinger operator as a true total Laplace-Beltrami operator
-- but defined for geometric structures going beyond the standard
understanding of metric, de Rham derivative etc. The approach is
again a bit similar to the idea of using a Lorenzian metric on the
extended space-time as present in \cite{DB}. However, this task
requires much more advanced geometrical techniques and we postpone
it to a separate paper as well as a generalization to curved
space-times.

\bigskip
\noindent Katarzyna Grabowska\\
Division of Mathematical Methods in Physics,
                University of Warsaw \\
                Ho\.za 69, 00-681 Warszawa, Poland \\
                 {\tt konieczn@fuw.edu.pl} \\\\
\noindent Janusz Grabowski\\Institute of Mathematics, Polish
Academy of Sciences\\ \'Sniadeckich 8, P.O. Box 21, 00-956
Warszawa,
Poland\\{\tt jagrab@impan.gov.pl}\\\\
\noindent Pawe\l\ Urba\'nski\\
Division of Mathematical Methods in Physics,
                University of Warsaw \\
                Ho\.za 69, 00-681 Warszawa, Poland \\
                 {\tt urba\'nski@fuw.edu.pl}

\end{document}